# GeoSES – UM ÍNDICE SOCIOECONÔMICO PARA ESTUDOS DE SAÚDE NO BRASIL

Ligia Vizeu Barrozo[1, 2, 3,*], Michel Fornaciali[2], Carmen Diva Saldiva de André[4], Guilherme Augusto Zimeo Morais[2], Giselle Mansur[1, 2], William Cabral-Miranda[1, 3], João Ricardo Sato[2], Edson Amaro Júnior[2]

[1]Universidade de São Paulo, Faculdade de Filosofia, Letras e Ciências Humanas, Departamento de Geografia, São Paulo, SP, Brasil
[2]Hospital Israelita Albert Einstein, Grupo de Análise de Big Data, São Paulo, SP, Brasil
[3]Universidade de São Paulo, Instituto de Estudos Avançados, São Paulo, SP, Brasil
[4]Universidade de São Paulo, Instituto de Matemática e Estatística, Departamento de Estatística, São Paulo, SP, Brasil

**Resumo**

Objetivo: definir um índice que resume as principais dimensões do contexto socioeconômico para fins de pesquisa, avaliação e monitoramento das desigualdades em saúde. Métodos: o índice foi criado a partir do Censo Demográfico de 2010, cuja seleção das variáveis foi orientada por referenciais teóricos para estudos em saúde, incluindo sete dimensões socioeconômicas: educação, mobilidade, pobreza, riqueza, renda, segregação e privação a recursos e serviços. O índice foi desenvolvido a partir de análise de componentes principais, sendo avaliado pelo seu *constructo, conteúdo e aplicabilidade*. Resultados: As dimensões do GeoSES-BR mostraram boa associação com o IDH-M (> 0,85). O modelo com a dimensão pobreza foi o que melhor explicou o risco relativo de mortalidade por causas evitáveis no Brasil. Na escala intraurbana, o modelo com o GeoSES-IM foi o que melhor explicou o risco relativo de mortalidade por doenças do aparelho circulatório. Conclusão: O GeoSES mostrou potencial explicativo significativo nas escalas estudadas.

**Abstract**

*Objective: to define an index that summarizes the main dimensions of the socioeconomic context for research purposes, evaluation and monitoring health inequalities. Methods: the index was created from the 2010 Brazilian Demographic Census, whose variables selection was guided by theoretical references for health studies, including seven socioeconomic dimensions: education, mobility, poverty, wealth, income, segregation and deprivation of resources and services. The index was developed using principal component analysis, and was evaluated for its construct, content and applicability components. Results: GeoSES-BR dimensions showed good association with HDI-M (above 0.85). The model with the poverty dimension best explained the relative risk of avoidable cause mortality in Brazil. In the intraurban scale, the model with GeoSES-IM was the one that best explained the relative risk of mortality from circulatory system diseases. Conclusion: GeoSES showed significant explanatory potential in the studied scales.*

# Introdução

O paradigma do Código de Endereçamento Postal (CEP) diz que o lugar onde o indivíduo mora é um preditor mais importante de sua saúde do que o seu próprio código genético.[1] De fato, no nível do indivíduo, a relação entre o status socioeconômico (SSE) e a prevalência de doenças crônicas apresenta um claro gradiente linear inverso.[2] Em outras palavras, à medida em que o SSE aumenta, a prevalência apresenta quedas compatíveis. Este gradiente é forte e consistente na relação encontrada com doenças cardiovasculares, diabetes tipo 2, síndrome metabólica, artrite, tuberculose respiratória crônica e desfechos adversos no nascimento, assim como com mortes violentas e acidentais.[2] No entanto, as condições individuais são insuficientes para explicar as causas das doenças e a relação entre desigualdades sociais e saúde. Estudos com dados agregados em áreas geográficas têm mostrado que as condições socioeconômicas dos lugares também afetam a saúde das pessoas.[3–5] Assim, compreender quais características do ambiente socioeconômico mais explicam as condições de saúde pode contribuir para a implantação de políticas públicas intersetoriais, que seriam mais eficientes para melhorar a saúde da população como um todo e reduzir as desigualdades.

Como medidas individuais de indicadores socioeconômicos raramente estão disponíveis nos registros médicos,[6] é comum a utilização de uma única variável geográfica que represente as condições de vida de forma resumida (e.g., renda, educação, PIB *per capita*). Embora esta abordagem auxilie na compreensão de como um aspecto do contexto socioeconômico está relacionado com a saúde, a interpretação dos achados fica limitada, já que o contexto socioeconômico é multidimensional, envolvendo aspectos, tais como, emprego, renda, educação, habitação, segregação, mobilidade, entre outros.[7] Por outro lado, ao se incluir mais de uma variável socioeconômica em um modelo de regressão — análise estatística comumente usada em estudos de delineamento ecológico — pode-se violar as premissas básicas desta análise devido aos efeitos de colinearidade.[8] A utilização de índices compostos pretende superar tais problemas, adicionando poder explicativo ao contexto socioeconômico dos lugares. Índices podem ser especialmente úteis se permitirem a avaliação de como determinada dimensão pode influenciar a saúde.[9]

No Brasil, estudos com abordagem ecológica em saúde costumam usar uma variável única como *proxy* do contexto socioeconômico — geralmente, a renda domiciliar mensal ou outras variáveis provenientes de estatísticas públicas. O Índice de Desenvolvimento Humano, calculado para os municípios (IDH-M),[10] é o indicador composto mais frequentemente utilizado em escala nacional. Sua principal vantagem, além de apresentar validade de constructo e confiabilidade, compreende a comparabilidade internacional. No entanto, este índice foi concebido para expressar o quanto o processo de desenvolvimento tem garantido oportunidades de "acesso à educação e cultura, a condições de desfrutar vida longa e saudável e a condições de dispor um padrão adequado de vida para a população".[11] Embora na prática tenha auxiliado a identificar desigualdades em desenvolvimento entre os municípios brasileiros, há redundância matemática quando é aplicado para explicar os desfechos de saúde. A componente longevidade do IDH compreende uma medida síntese das condições de saúde e riscos à morbimortalidade. Há, portanto, carência de um índice sintético das condições socioeconômicas elaborado a partir de referenciais teóricos para os estudos de saúde no Brasil.

Com o objetivo de suprir tal carência, introduzimos o **GeoSES** (Índice Socioeconômico do Contexto Geográfico para Estudos em Saúde). O GeoSES sintetiza as dimensões socioeconômicas mais relevantes para contextualizar a saúde, visando fins de pesquisa, avaliar e monitorar desigualdades, e desenvolver estratégias de alocação de recursos e serviços. O novo índice foi desenvolvido para utilização em três escalas de agregação: nacional (GeoSES-BR), estadual (GeoSES-UF) e intramunicipal (GeoSES-IM, para municípios com três ou mais áreas de ponderação). Aqui apresentamos a aplicação nas escalas nacional e intramunicipal.

# Métodos

## Unidades geográficas

O setor censitário é a menor unidade geográfica para a qual o Censo Demográfico decenal brasileiro tabula dados do universo dos domicílios. Uma segunda unidade censitária corresponde à *área de ponderação*, que contém um grupo de setores censitários. Nesta unidade são aplicados procedimentos estatísticos que permitem levantamentos amostrais válidos para toda a população. Esta amostra responde a um questionário que inclui quesitos do questionário básico universal, somados a outros de investigação mais detalhada sobre características do domicílio e de seus moradores.[12] Assim, o índice sintético foi criado a partir da seleção das variáveis contidas no questionário da Amostra do Censo Demográfico de 2010.

## Variáveis socioeconômicas e dimensões do contexto

A escolha das variáveis foi orientada por conceitos teóricos para estudos em saúde,[7,13] seguindo agrupamento de sete dimensões do contexto socioeconômico. As dimensões (e seu total de variáveis), são: educação (7), mobilidade (6), pobreza (5), riqueza (3), renda (1), segregação (5) e privação a recursos e serviços (19). As 46 variáveis utilizadas, e seus significados, estão listadas no Anexo I. As dimensões *renda* e *educação* podem influenciar a etiologia de vários desfechos em saúde, em parte, através de mecanismos que envolvem a obtenção de recursos materiais. A *educação* pode refletir o conhecimento geral e o sanitário, que podem ser independentes da renda, capacidade de solucionar problemas, prestígio, influência, participação em redes sociais e acesso às inovações tecnológicas,[14] que podem trazer vantagens para a saúde de um indivíduo.[9]

A dimensão *pobreza* refere-se à pobreza absoluta, diretamente ligada à capacidade mínima de sobrevivência e acesso a recursos materiais. A *riqueza*, por sua vez, difere da dimensão renda, pois é uma *proxy* para todos os recursos econômicos que foram acumulados ao longo da vida.[9] A dimensão *privação a recursos e acesso a serviços públicos* se sustenta no conceito de *privação material* de Tonwsend,[15] que diz respeito à desvantagem em relação à sociedade à qual o indivíduo pertence. Aqui pretende-se medir o quanto uma pessoa dispõe de recursos materiais e conveniências que fazem parte da vida moderna, tais como habitação adequada, posse de carro, geladeira, computador, entre outros; e o acesso a serviços, dentre eles, saneamento, energia elétrica, e internet.

A *mobilidade* pode afetar a saúde de uma pessoa em diversos aspectos. Um deles diz respeito ao tempo dispendido no deslocamento da casa ao trabalho, que pode causar estresse em vários níveis, além de comprometer o tempo disponível para estudo ou lazer. Além disso, o maior tempo no deslocamento pode expor as pessoas a maiores doses de poluição do ar nas grandes cidades.[16]

Por fim, a *segregação residencial* é um conceito amplo que se refere à habitação separada de diferentes grupos populacionais em diferentes partes de uma cidade.[17] A segregação afeta a saúde por intensificar efeitos psicossociais que envolvem insegurança, ansiedade, isolamento social, ambientes socialmente perigosos, *bullying* e depressão.[18–20]

### Obtenção e processamento dos dados

Os dados utilizados para geração do índice são provenientes do Censo 2010, disponibilizados pelo IBGE por meio do seu serviço de FTP (*File Transfer Protocol*), organizados por Unidade Federativa (UF), seguidamente agrupados por informações sobre "Pessoas" e "Domicílios". Os dados foram filtrados de acordo com os referenciais teóricos[13,21] relacionados ao interesse do estudo. Os dados filtrados foram armazenados em arquivos .csv, os quais foram processados novamente para melhor organização e facilidade de manipulação automática.

Em seguida, os arquivos resultantes do passo anterior foram processados para agregar as informações originais do Censo gerando as variáveis de interesse da metodologia. Este passo foi composto por duas atividades distintas:

- A maioria das variáveis originais do Censo foram processadas para gerar valores em porcentagem relativos ao total da população de uma região de interesse, considerando o peso amostral da área de ponderação. Por exemplo, ao quantificar pessoas de um município que possuem ensino superior completo, o resultado do processamento é um valor que representa a porcentagem (de 0 a 100) daquela população que possui ensino superior completo. Exceções se aplicam às variáveis de significado direto, como renda, nas quais foram utilizados os valores médios, considerando o peso amostral, resultantes da ponderação dos valores originais pelas regiões de agregação (área de ponderação ou município);
- As informações processadas no item anterior foram agrupadas considerando os pesos amostrais, e disponibilizadas de acordo com a abrangência do índice a ser gerado, ou seja, para uma abrangência municipal as informações foram disponibilizadas por áreas de ponderação; para uma abrangência estadual as informações foram disponibilizadas, de modo consolidado, para cada município daquele Estado;

Ao final deste processo, tem-se as informações no formato esperado para cálculo do índice, separadas por UF e tipo (área de ponderação ou município).

### Definição do GeoSES

O índice foi desenvolvido aplicando-se de forma sucessiva a técnica de análise de componentes principais (ACP)[22], com base na metodologia de Lalloué *et al.*,[23] considerando a matriz de correlação das variáveis originais. As componentes principais são combinações lineares das variáveis originais, construídas de tal forma que sejam independentes entre si, e a primeira tenha variância máxima, a segunda a maior variância entre as demais e assim sucessivamente. O número de componentes que podem ser construídas é igual ao número de variáveis no estudo, mas, de uma forma geral, com um número menor de componentes é possível explicar praticamente toda a variabilidade dos dados, ou seja, há redução de dimensão. Foram utilizados os dados descritos anteriormente, selecionados de acordo com à área de interesse sob análise, a qual pode ser: nacional (o País todo, utilizando dados consolidados de todos os municípios), estadual (indicador de cada Estado, utilizando dados de seus municípios) ou intramunicipal (para um município específico, utilizando dados das áreas de ponderação). Os passos para geração do índice são os mesmos independentemente da área de interesse. Um projeto no Município de São Paulo foi desenvolvido inicialmente [24] para posterior aplicação em escala nacional.

PRÉ-PROCESSAMENTO:

Iniciou-se o cálculo do índice pela leitura dos dados descritos na seção anterior. Em seguida, foi somada uma constante igual a 10 a todos os valores lidos, pois muitas das variáveis são expressas em intervalo de 0 a 100, sendo que o valor 0 (zero) pode gerar problemas em operações numéricas. Esta mudança de escala não interfere nos resultados porque não afeta a estrutura de correlação dos dados

PASSOS:
1) Foi realizada uma ACP em cada dimensão. O número de componentes selecionadas foi tal que a porcentagem da variância total explicada fosse maior ou igual a 75%. Por facilidade de interpretação, foram consideradas as variáveis com o maior coeficiente em cada componente.
2) Considerando as variáveis selecionadas no passo 1), foi feita uma ACP e considerada a primeira componente. As variáveis cujos valores absolutos dos coeficientes ficaram acima da média dos coeficientes foram selecionadas.
3) Foi aplicada a técnica de ACP às variáveis selecionadas no passo 2). A primeira componente principal resultante consiste no GeoSES (Índice Socioeconômico do Contexto Geográfico para Estudos em Saúde).
4) Foram então calculados os valores (escores) do GeoSES.
5) Os escores foram padronizados para o intervalo [-1,+1].

INTERPRETAÇÃO:
Assim como o IDH é decomposto em Longevidade (IDH-L), Renda (IDH-R) e Educação (IDH-E), foi desenvolvida uma forma de expressar a contribuição de cada dimensão do GeoSES para melhor interpretabilidade dos resultados. No caso de haver mais de uma variável em uma mesma dimensão, foi representada aquela que foi mais correlacionada com o GeoSES. Entende-se que ao selecionar apenas a variável mais correlacionada, utiliza-se a medida mais representativa daquela dimensão. Isto evita, ao mesmo tempo, processamentos complexos para, eventualmente, considerar mais de uma variável por dimensão, o que poderia levar a resultados não interpretáveis.

### Criação do GeoSES

Uma vez definida a metodologia descrita acima e validada para o município de São Paulo/SP, o processo foi implementado em linguagem de desenvolvimento Python, permitindo a escalabilidade da geração do índice para todo o território nacional, garantindo agilidade e consistência dos resultados.

Para o desenvolvimento computacional, adotou-se a parametrização do método de forma modularizada. Com isso, ao explicitar quais dimensões são utilizadas e quais variáveis estão contempladas em cada dimensão, facilita-se a aplicabilidade da metodologia para dados de outros Censos, tanto passados quanto futuros.

Para facilitar o uso e disseminação dos resultados, todos os índices e suas informações associadas são disponibilizados em mapas interativos no formato HTML (veja um exemplo no Anexo II). Cada município ou estado possui um arquivo com seu nome no qual é possível observar, interativamente, a distribuição geográfica dos dados. Há uma camada no mapa para cada dimensão utilizada na análise, além de uma camada principal que ilustra a distribuição espacial do índice em si.

## Resultados

A avaliação do GeoSES compreendeu seu *conteúdo*, seu *constructo* e sua *aplicabilidade* para estudos na saúde. A validação de *conteúdo* verifica a relevância e representatividade das dimensões que compõem o GeoSES para descrever o fenômeno medido[25]. Devido ao mecanismo da ACP, as variáveis selecionadas são naturalmente representativas das dimensões às quais pertencem, e as dimensões que compõem o GeoSES são relevantes para contextualizar os fenômenos socioeconômicos devido aos referenciais teóricos. Além disso, calculamos o coeficiente alfa de Cronbach para os índices gerados, obtendo os valores de 0,93, 0,89 e 0,97 para o GeoSES-BR, GeoSES-UF e GeoSES-IM, respectivamente. Lembramos que quanto mais próximo de 1, mais homogêneas são as variáveis do índice.

### Avaliação do constructo do GeoSES

A validação do *constructo* verifica, do ponto de vista teórico, se o novo índice está associado ao conceito que supostamente deve medir. Avaliamos o GeoSES comparando-o com o IDH-M: um indicador amplamente aceito e aplicado para o mesmo nível de agregação (no caso, para municípios). A associação do GeoSES com o IDH-M foi realizada de modo qualitativo e quantitativo.

Em termos qualitativos, a média do GeoSES-BR entre os municípios do Brasil foi -0,40. Melgaço/PA recebeu a pior avaliação (-1) e Santana de Parnaíba/SP, a melhor (+1). Segundo o IDH-M, o IDH médio dos municípios brasileiros é 0,659. A pior posição também é ocupada por Melgaço/PA (0,418), mas a melhor, por São Caetano do Sul/SP (0,862). Tal comparação reforça a similaridade entre os índices, porém destaca que diferenças podem surgir, as quais potencialmente melhor explicam as condições socioeconômicos das regiões brasileiras.

Em termos quantitativos, a Tabela 1 mostra as correlações entre "GeoSES *vs.* IDH-M" e suas "Dimensões *vs.* Componentes". As dimensões do GeoSES mostraram boa correlação com o IDH-M, acima de 0,85, exceto riqueza e segregação.

**Tabela 1** – Matriz de correlação entre os índices e suas dimensões na escala nacional, agregação por município.

|         | GeoSES | educ  | pobreza | privação | riqueza | renda | segreg | IDHM | IDHM_E | IDHM_L | IDHM_R |
|---------|--------|-------|---------|----------|---------|-------|--------|------|--------|--------|--------|
| **GeoSES**  | 1      |       |         |          |         |       |        |      |        |        |        |
| **educ**    | -0.86  | 1     |         |          |         |       |        |      |        |        |        |
| **pobreza** | -0.96  | 0.81  | 1       |          |         |       |        |      |        |        |        |
| **privação**| 0.93   | -0.74 | -0.90   | 1        |         |       |        |      |        |        |        |
| **riqueza** | 0.57   | -0.61 | -0.43   | 0.41     | 1       |       |        |      |        |        |        |
| **renda**   | 0.93   | -0.82 | -0.88   | 0.83     | 0.60    | 1     |        |      |        |        |        |
| **segreg**  | 0.82   | -0.54 | -0.77   | 0.80     | 0.27    | 0.67  | 1      |      |        |        |        |
| **IDHM**    | 0.94   | -0.93 | -0.93   | 0.86     | 0.53    | 0.89  | 0.70   | 1    |        |        |        |
| **IDHM_E**  | 0.85   | -0.95 | -0.82   | 0.76     | 0.51    | 0.78  | 0.60   | 0.95 | 1      |        |        |
| **IDHM_L**  | 0.82   | -0.71 | -0.83   | 0.78     | 0.41    | 0.76  | 0.64   | 0.85 | 0.70   | 1      |        |
| **IDHM_R**  | 0.95   | -0.84 | -0.96   | 0.87     | 0.53    | 0.94  | 0.73   | 0.95 | 0.82   | 0.83   | 1      |

## Avaliação da aplicabilidade do GeoSES na saúde

Como estágio adicional de avaliação, validamos o potencial explanatório do GeoSES para desfechos de saúde em duas escalas de agregação: *nacional* (GeoSES-BR), utilizando o cálculo do risco relativo para óbitos por causas evitáveis de 5 a 74 anos por intervenções do Sistema Único de Saúde do Brasil, entre 2013 e 2017[30]; *intramunicipal* (GeoSES-IM), utilizando o cálculo do risco relativo de mortalidade por doenças do aparelho circulatório dos residentes de São Paulo/SP, por área de ponderação, entre 2006 e 2009.

A motivação para escolha de tais desfechos e períodos resulta da disponibilidade dos dados, os quais foram padronizados por sexo e faixa etária detalhada. Os dados foram obtidos, respectivamente, das "Estatísticas Vitais" e do "Sistema de Informações sobre Mortalidade" (SIM) do DATASUS. Em particular, óbitos relacionados às doenças do aparelho circulatório são um desfecho reconhecidamente associado com as condições socioeconômicas em nível agregado. Seus dados foram geocodificados por área de ponderação pela Fundação SEADE.

Utilizamos modelos de regressão linear simples para validar o poder explicativo do GeoSES na saúde. Comparamos a regressão dos desfechos com o GeoSES (e suas dimensões), contra a regressão dos desfechos com o IDH-M (e seus componentes).

Iniciou-se com a verificação dos pressupostos para a regressão de Mínimos Quadrados Ordinários entre os desfechos e os índices estudados e a análise de dependência espacial dos resíduos. Utilizamos as coordenadas geográficas das sedes municipais, nas análises para o Brasil e as coordenadas deslocadas das áreas de ponderação para as análises intraurbanas de São Paulo/SP. Neste caso, o deslocamento foi feito devido à heterogeneidade da distribuição da população nas áreas de ponderação periféricas, onde há represas e áreas de proteção ambiental. Devido à dependência espacial nos resíduos nas duas escalas de agregação, aplicamos modelos de regressão geograficamente ponderados, calculados no programa ArcGIS 10.1 (análise Kernel do tipo "*adaptive*"; método de largura de banda *Bandwidth Parameter*, com 53 vizinhos na escala nacional, e 30, na intramunicipal). Avaliamos os modelos via AIC (*Akaike Information Criterion*) da regressão, segundo o qual, menores valores indicam melhor ajuste do modelo. Para verificar a dependência espacial, calculamos os coeficientes I de Moran para os resíduos padronizados e seus valores *p* no programa GeoDa com matriz de vizinhança do tipo "Queen".

Os resultados apontam que, na escala nacional, o modelo com o GeoSES apresentou melhor ajuste (AIC: -4583,44), quando comparado com aquele em que foi considerado o IDH-M (AIC: -524,22), embora ambos tenham apresentado dependência espacial em seus resíduos (Tabela 2). As dimensões pobreza, privação, renda e segregação foram as que melhor explicaram o risco relativo de mortalidade por causas evitáveis de 5 a 74 anos no Brasil, sem dependência espacial em seus resíduos. Entre as dimensões deste índice, a que mais explica a variabilidade espacial do risco é a pobreza (Tabela 2). O mapa da Figura 1 mostra os riscos observados e os riscos explicados pelo GeoSES-BR/pobreza. A dimensão mobilidade não foi um critério determinístico para caracterizar diferenças socioeconômicas em escala nacional e não contribuiu para o GeoSES-BR.

**Tabela 2** – Resultados dos modelos de regressão simples geograficamente ponderados (RGP) entre risco relativo de mortalidade por causas evitáveis de 5 a 74 anos padronizado e índices (IDH-M e GeoSES-BR) e suas dimensões – valores de R2 global ajustado, Critério de Informação de Akaike (AIC), coeficientes I de Moran e valor p para dependência espacial.

| Indicador | $R^2$ global ajustado | AIC | Coeficiente I de Moran | Valor p |
|---|---|---|---|---|
| **IDH-M** | 0,504 | -524,22 | 0,016 | 0,014# |
| **IDHM-E** | 0,406 | -3817,45 | 0,017 | 0,008# |
| **IDHM-L** | -- | -- | -- | -- |
| **IDHM-R** | 0,529 | -1017,92 | 0,017 | 0,010# |
| **GeoSES-BR** | 0,422 | -4583,44 | 0,015 | 0,029# |
| **GeoSES-renda** | 0,428 | -4636,58 | 0,009 | 0,107 |
| **GeoSES-educação** | 0,434 | -3290,28 | 0,015 | 0,013# |
| **GeoSES-riqueza** | 0,398 | -4344,18 | 0,022 | 0,002# |
| **GeoSES-privação** | 0,428 | -4648,99 | 0,000 | 0,465 |
| **GeoSES-segregação** | 0,380 | -4073,14 | 0,011 | 0,066 |
| **GeoSES-pobreza** | 0,434 | -4702,75* | 0,012 | 0,067 |

\* melhor ajuste; #dependência espacial nos resíduos; -- multicolinearidade local não permitiu o cálculo do modelo

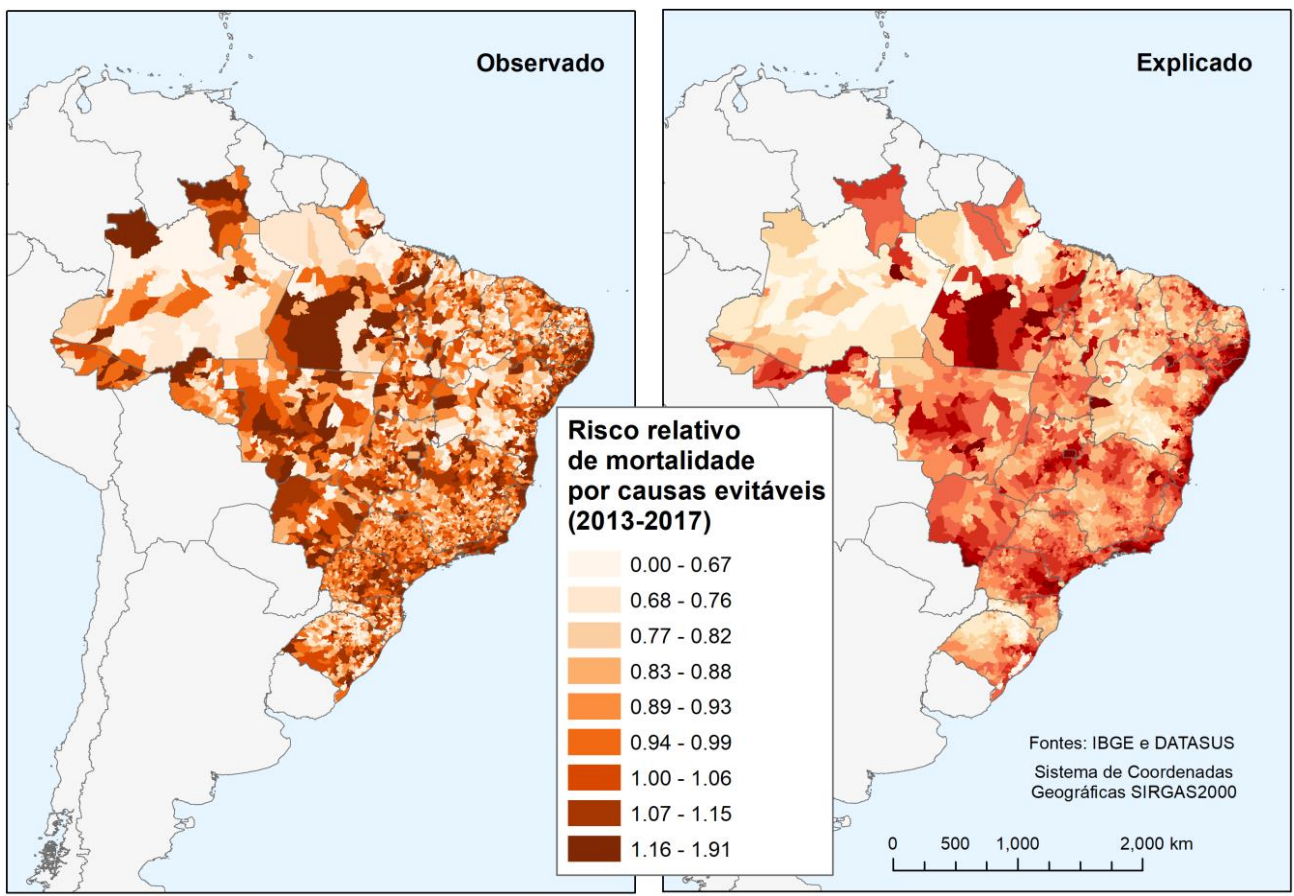

**Figura 1** – Riscos relativos de mortalidade por causas evitáveis de 5 a 74 anos (2013 a 2017): risco relativo observado e risco relativo explicado pelo GeoSES-pobreza.

Já na escala intramunicipal, o GeoSES-IM explica cerca de 67% da variabilidade do desfecho (AIC: -357,86). Neste caso a síntese das dimensões trouxe contribuição evidente, já que nenhuma dimensão isoladamente teve desempenho superior ao GeoSES-IM (Tabela 3). A Figura 2 compara os valores observados e explicados.

**Tabela 3** – Resultados dos modelos de regressão simples geograficamente ponderados (RGP) entre risco relativo de mortalidade por doenças do aparelho circulatório no município de São Paulo e o índice GeoSES-IM e suas dimensões – valores de R2 global ajustado, Critério de Informação de Akaike (AIC), coeficientes I de Moran e valor p para dependência espacial.

| Indicador | R2 global ajustado | AIC | Coeficiente I de Moran | Valor p |
|---|---|---|---|---|
| GeoSES-IM | 0,673 | -357,86* | -0,032 | 0,196 |
| GeoSES-renda | 0,644 | -333,42 | -0,020 | 0,331 |
| GeoSES-educação | 0,649 | -338,72 | -0,029 | 0,213 |
| GeoSES-riqueza | 0,618 | -313,86 | -0,012 | 0,436 |
| GeoSES-privação | 0,594 | -297,72 | -0,037 | 0,146 |
| GeoSES-segregação | 0,651 | -338,15 | -0,023 | 0,298 |
| GeoSES-pobreza | 0,628 | -313,28 | -0,041 | 0,117 |
| GeoSES-mobilidade | 0,574 | -276,01 | -0,026 | 0,261 |

\* melhor ajuste

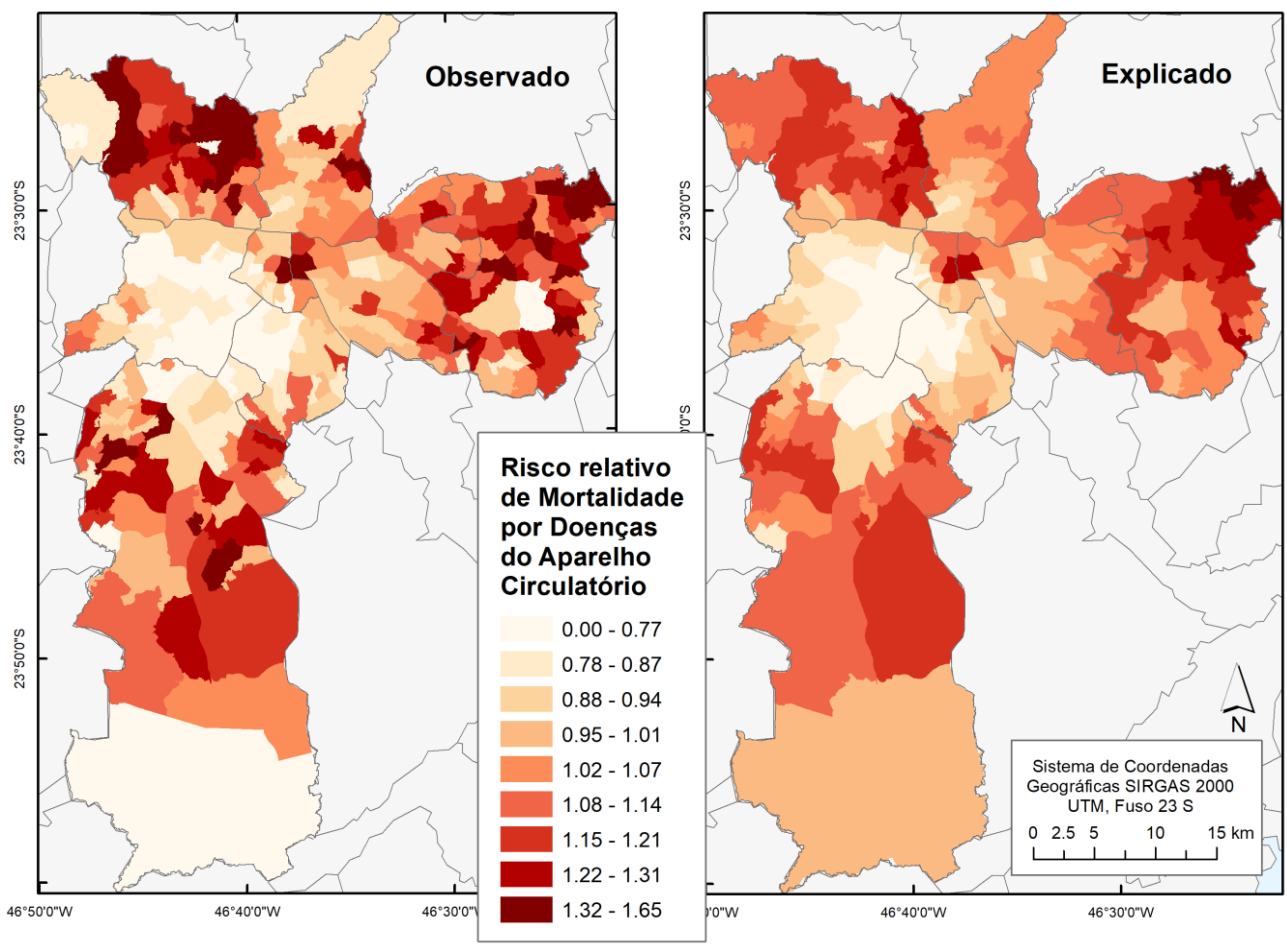

**Figura 2** – Riscos relativos de mortalidade por doenças do aparelho circulatório (2006 a 2009) no Município de São Paulo (SP): risco relativo observado e risco relativo explicado pelo GeoSES.

Portanto, o GeoSES mostrou potencial explicativo significativo nas duas escalas estudadas. Como a saúde tem causas múltiplas, não se espera que apenas o GeoSES esclareça toda a variabilidade espacial de um desfecho. Outros aspectos contribuem na compreensão dos fenômenos, tais como condições do ambiente natural e construído, acesso e qualidade dos serviços à saúde, e questões políticas e macroeconômicas.

## Discussão

A relação entre as condições socioeconômicas e a saúde, apesar de ser um tema da origem da Epidemiologia Social, ainda diverge quanto à definição de seus indicadores e quanto à contribuição que exercem[26–28]. O consenso atual reafirma a complexidade do tema ao constatar que não existe um indicador universal capaz de explicar todos os desfechos.[9] Assim, é comum a escolha de uma variável unidimensional para representá-lo, incorrendo em simplificações que não contribuem para a compreensão do problema estudado. Também é trivial a utilização de índices que não foram elaborados para esta finalidade e que não permitem identificar os aspectos mais importantes para a saúde.

Neste estudo, apresentamos um índice capaz de sintetizar sete dimensões que compõem o contexto socioeconômico, permitindo uma avaliação global do contexto e suas particularidades a partir de diferentes aspectos. O GeoSES não pretende explicar todos os determinantes da saúde, mas permite identificar se o contexto socioeconômico se relaciona ao desfecho estudado, o quanto e quais aspectos se destacam. Em particular, a segregação residencial no Brasil — embora conceitualmente relevante — tem sido pouco avaliada em estudos de saúde. [29] Principalmente em escala intramunicipal, a segregação residencial se reflete em muitos desfechos. Evidenciá-la pode incentivar políticas afirmativas de inclusão.

Como o GeoSES foi implementado em linguagem de programação, ele poderá ser facilmente atualizado a cada edição do Censo. Também pode ser adaptado aos Censos anteriores. Além disso, outras versões do GeoSES podem ser desenvolvidas, permitindo seu aprimoramento ao incluir dimensões que não foram contempladas nesta primeira versão.

Destaca-se, no entanto, que há uma correlação significativa do novo índice com o IDH-M, pois o conjunto de dados utilizados para compô-los é semelhante. Contudo, tem-se a vantagem de poder decompor o novo índice em até sete dimensões, ao contrário do IDH-M, que utiliza apenas três. Ainda, a geração do índice em nível intraurbano permite aos gestores de Saúde tomar decisões em nível municipal por meio da observação das semelhanças/diferenças entre regiões de uma mesma cidade.

A elaboração do GeoSES parte de referenciais teóricos, mas permite que as variáveis mais explicativas matematicamente sejam escolhidas pela análise estatística. Neste sentido, a variável mais explicativa não é arbitrária, mas a mais discriminante, podendo ser diferente de acordo com a região sob análise (nacional, estadual e municipal). Isso traz uma perspectiva inovadora e rica para a compreensão do que é mais relevante no contexto socioeconômico de cada região brasileira. Por exemplo, a dimensão mais significante em nível nacional é a pobreza; já para o estado de São Paulo, é a segregação residencial do contexto educacional; para o estado do Rio Grande do Norte, a renda é a dimensão mais explicativa. Na escala intramunicipal das capitais das UFs, as diversas formas de segregação residencial aparecem como importante variável discriminante. Por este motivo, este índice pode ser usado em estudos de outras áreas do conhecimento, tais como Geografia, Sociologia e Economia.

Os resultados disponibilizados[1] — mapas interativos e valores tabulados por região de interesse — poderão contribuir para ações futuras relevantes. No âmbito científico, o índice pode embasar estudos dos aspectos específicos das desigualdades em saúde e dos mecanismos que as conduzem. No âmbito prático, o índice poderá orientar a elaboração de políticas públicas intersetoriais ou nas gestões estaduais e municipais.

# Agradecimento

# Contribuição dos Autores

LVB, CDSA, JRS e EAJ delinearam o estudo; LVB, CDSA, WCM e MSF desenvolveram o índice; MSF e GAZM implementaram o índice em linguagem computacional e visualização HTML; LVB, WCM, e GMB conduziram as regressões espaciais. Todos os autores contribuíram na interpretação dos resultados e redação do artigo, leram e revisaram a última versão.

# Referências

---

[1] Os mapas interativos e os valores tabulados por região de interesse serão disponibilizados pelo Ministério da Saúde em local a ser definido em breve.

# Anexo I

Variáveis utilizadas para composição do GeoSES.

| VARIÁVEL | SIGNIFICADO |
|---|---|
| **Dimensão "Educação"** | |
| **P_GRAD** | porcentagem de pessoas cuja espécie do curso mais elevado concluído foi graduação |
| **P_MEST** | porcentagem de pessoas cuja espécie do curso mais elevado concluído foi mestrado |
| **P_DOUTOR** | porcentagem de pessoas cuja espécie do curso mais elevado concluído foi doutorado |
| **P_SEM_INST** | porcentagem de pessoas cujo nível de instrução é o sem instrução e fundamental incompleto |
| **P_FUND** | porcentagem de pessoas cujo nível de instrução é o fundamental completo e médio incompleto |
| **P_ENSMED** | porcentagem de pessoas cujo nível de instrução é o Médio completo e superior incompleto |
| **P_ENSSUP** | porcentagem de pessoas cujo nível de instrução é o superior completo |
| **Dimensão "Mobilidade"** | |
| **P_OUTROMUNC** | porcentagem de pessoas que trabalha em outro município |
| **P_CASADIA** | porcentagem de pessoas que retorna diariamente do trabalho para casa |
| **P_ATE5** | porcentagem de pessoas cujo tempo habitual gasto de deslocamento de sua casa até o trabalho é de até 5 minutos |
| **P_6A30** | porcentagem de pessoas cujo tempo habitual gasto de deslocamento de sua casa até o trabalho é de até 6 a 30 minutos |
| **P_1A2** | porcentagem de pessoas cujo tempo habitual gasto de deslocamento de sua casa até o trabalho é de 1 a 2 horas |
| **P_MAISDE2** | porcentagem de pessoas cujo tempo habitual gasto de deslocamento de sua casa até o trabalho é de mais de 2 horas |
| **Dimensão "Pobreza"** | |
| **MEDIA_DENSMORA** | densidade de morador por cômodo |
| **P_POBREZA** | % de pessoas na linha da pobreza: cujo rendimento domiciliar per capita é menor ou igual a R$255,00 (meio salário mínimo em 2010) |
| **P_PPI_POBREZA** | % de pessoas na linha da pobreza e de raça ou etnia preta, parda ou indígena |
| **P_BOLSA_FAM** | porcentagem de pessoas que em julho de 2010, tinham rendimento mensal habitual de Programa Social Bolsa-Família ou Programa de Erradicação do Trabalho Infantil (PETI): |
| **P_OUTROSPROG** | porcentagem de pessoas que em julho de 2010, tinham rendimento mensal habitual de outros programas sociais ou de transferências |
| **Dimensão "Privação material e social"** | |
| **P_ALVSREV** | porcentagem de domicílios de alvenaria sem revestimento |
| **P_REDE_ESG** | porcentagem de domicílios com rede geral de esgoto |
| **P_REDE_AGUA** | porcentagem de domicílios com rede geral de distribuição de água |
| **P_LIXO** | porcentagem de domicílios com lixo coletado diretamente por serviço de limpeza |
| **P_ENERGIA** | porcentagem de domicílios com energia elétrica de companhia distribuidora de energia |
| **P_TV** | porcentagem de domicílios com existência de TV |
| **P_MAQLAV** | porcentagem de domicílios com existência de máquina de lavar roupa |
| **P_GELADEIRA** | porcentagem de domicílios com existência de geladeira |
| **P_MAQTVGEL** | porcentagem de domicílios com existência de máquina de lavar, TV e geladeira |
| **P_CELULAR** | porcentagem de domicílios com existência de celular |
| **P_COMP_INT** | porcentagem de domicílios com existência de computador com acesso à internet |
| **P_CELCOMPINT** | porcentagem de domicílios com existência de telefone celular e computador com internet |
| **P_MOTO** | porcentagem de domicílios com existência de motocicleta para uso particular |
| **P_CARRO** | porcentagem de domicílios com existência de automóvel para uso particular |
| **P_ADEQ** | porcentagem de domicílios com moradia adequada |
| **P_TUDOADEQ** | porcentagem de domicílios com acesso a rede de esgoto, rede de água, coleta de lixo, energia elétrica e moradia adequada |
| **P_NEM_MOTO_CARRO** | porcentagem de domicílios sem moto ou carro para uso particular |
| **P_SO_MOTO** | porcentagem de domicílios com existência de apenas moto para uso particular |
| **P_SO_CARRO** | porcentagem de domicílios com existência de apenas carro para uso particular |
| **Dimensão "Renda"** | |
| **MED_RENDDOM** | rendimento mensal domiciliar em julho de 2010 |
| **Dimensão "Riqueza"** | |
| **P_ALUG1000** | porcentagem de domicílios alugados com valor de aluguel de R$1000,00 ou mais |
| **P_BANH4OUMAIS** | porcentagem de domicílios com 4 banheiros ou mais |
| **P_IDOSO10SM** | % de pessoas de 65 anos ou mais com rendimento mensal igual ou acima de R$5100,00 (ou 10 salários mínimos) |
| **Dimensão "Segregação"** | |
| **ICE_renda** | (número de pessoas com renda acima de R$5400,00-número de pessoas com renda abaixo de |
| | R$1000,00)/número de pessoas que responderam [os valores foram calculados com base nos percentis 20 e 80 do rendimento V6529 da planilha PESSOA dos microdados do Censo de 2010] |
| **ICEedu** | (número de pessoas com ensino superior completo- número de pessoas sem instrução e fundamental incompleto)/total de pessoas que responderam [V6400] |
| **ICE_renda_preto** | (número de brancos com rendimento acima de R$5400,00-número de pretos com rendimento igual ou menor do que R$1000,00)/total de pessoas que responderam as duas perguntas [V6529 e V0606] |
| **ICE_renda_ppi** | (número de brancos com rendimento acima de R$5400,00-número de pretos+pardos + indígenas com rendimento igual ou menor do que R$1000,00)/total de pessoas que responderam as duas perguntas [V6529 e V0606] |
| **ICE_branco_renda** | (número de brancos com rendimento acima de R$5400,00-número de brancos com rendimento igual ou menor do que R$1000,00)/total de pessoas que responderam as duas perguntas [V6529 e V0606] |

# Anexo II

Mapa interativo do GeoSES-SP. A figura mostra a distribuição geográfica dos valores do GeoSES, considerando o estado de São Paulo. Em destaque, o município de São Paulo/SP apresenta seu índice e os valores de suas dimensões. Nesta análise, nota-se que a dimensão "mobilidade" não é ativada, ou seja, não é significativa para caracterizar as diferenças socioeconômicas do estado. Além da camada principal (o índice), também é possível plotar cada uma das dimensões significativas da análise.

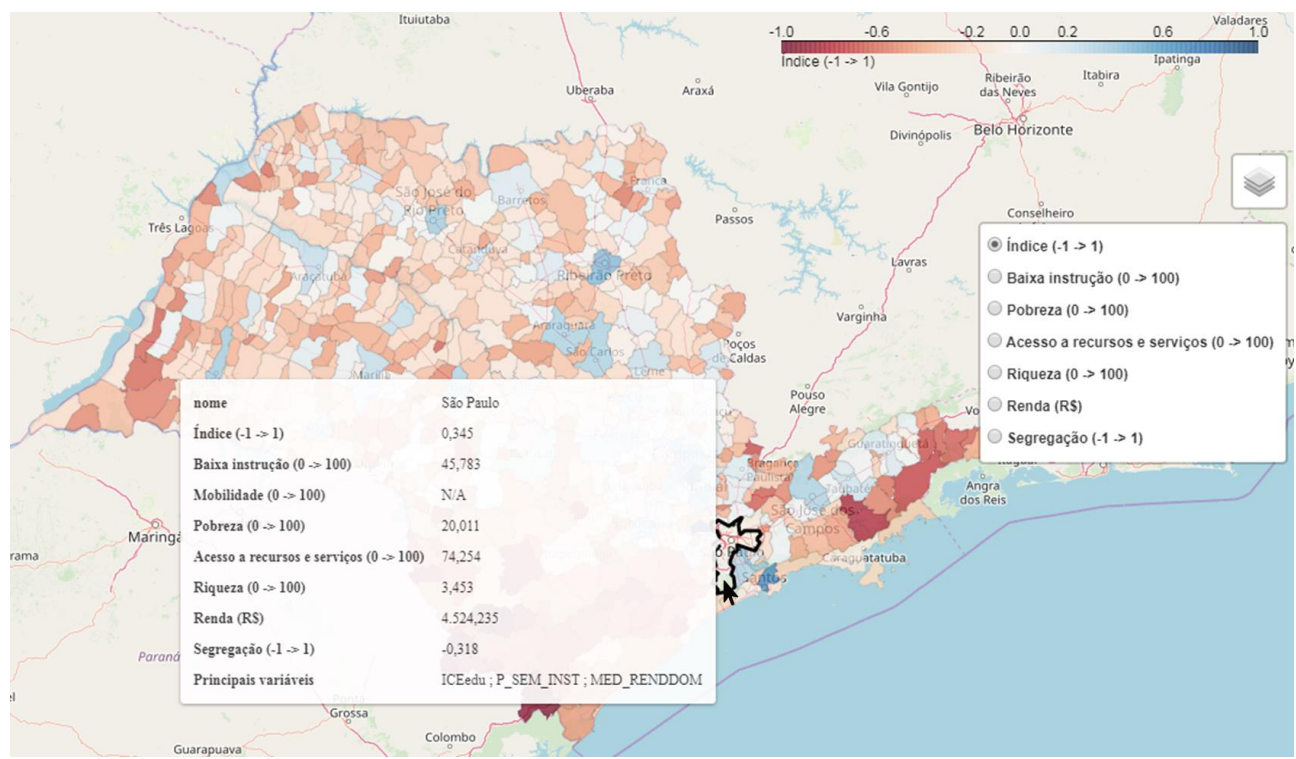